\documentclass[aps,prd,showpacs,superscriptaddress,nofootinbib]{revtex4}
\usepackage{graphicx}
\usepackage{graphics}
\usepackage{dcolumn}
\usepackage{amssymb}
\usepackage{bm}
\usepackage{amsmath}
\usepackage{color}
\usepackage{enumerate}

\newcommand{\fracc}[2]{\frac{\textstyle{#1}}{\textstyle{#2}}}

\begin{document}

\title{On the initial singularity problem in rainbow cosmology}
\author{Grasiele Santos}
\affiliation{Dipartimento di Fisica, Universit\`a ``La Sapienza'', P.le A. Moro 2, 00185 Roma, Italia
and CAPES Foundation, Ministry of Education of Brazil, Brasilia, Brazil}

\author{Giulia Gubitosi}
\affiliation{Theoretical Physics, Blackett Laboratory, Imperial College, London, SW7 2BZ, United Kingdom}

\author{Giovanni Amelino-Camelia}
\affiliation{Dipartimento di Fisica, Universit\`a ``La Sapienza'' and Sez. Roma1 INFN, P.le A. Moro 2,
00185 Roma, Italia}

\date{\today}

\begin{abstract}

It has been recently claimed that the initial singularity might be avoided
in the context of rainbow cosmology, where one attempts to account for  quantum-gravitational
corrections through an effective-theory description based
on an energy-dependent (``rainbow") space-time metric. We here scrutinize this exciting hypothesis
much more in depth than previous analyses.
In particular, we take into account all requirements for singularity avoidance,
while previously only a subset of these requirements had been considered.
Moreover, we show that the implications of a rainbow metric for thermodynamics
are more significant than previously appreciated.
Through the analysis of two particularly meaningful examples of rainbow metrics
we find that our concerns are not merely important conceptually, but actually
change in quantitatively significant manner the outcome of the analysis. Notably we only find examples where the singularity is not avoided, though one can have that in the regime where our semi-classical picture is still reliable the approach to the singularity is slowed down when compared to the standard classical scenario. We conclude that the study of rainbow metrics provides tantalizing hints of singularity avoidance but is inconclusive, since some key questions remain to be addressed just when the scale factor is very small, a regime which, as here argued, cannot be reliably described by an effective rainbow-metric picture.

\end{abstract}


\maketitle

\section{Introduction}
The avoidance of the initial singularity has been subject of intense debate in
cosmology (see Ref.\cite{novello} and references therein). In particular, since we
expect that general relativity (GR) should be no longer reliable under the extreme conditions of the primordial universe,
where the curvature and/or its invariants diverge, quantum gravitational effects might provide a
mechanism for avoiding the singularity. Examples of some attempts along this direction are given in Ref.\cite{asht11}.

In order to investigate this possibility one of the most natural and manageable starting points
is provided by the ``rainbow gravity" picture. There one assumes that quantum gravity
should admit a semi-classical regime in
which quantum-gravitational corrections can be effectively described by introducing
an energy-dependent spacetime metric, while otherwise keeping the framework rather conventional,
with spacetime geometry coded in a smooth manifold. This was inspired by (and provides a link with)
the intense effort devoted to the study \cite{AmelinoCamelia:1997gz, Gambini:1998it, Alfaro:1999wd, AmelinoCamelia:1999pm}
of the hypothesis of Planck-scale-modified
dispersion relations (MDR): with an energy-dependent spacetime metric
particles of different energies would experience different geometries of spacetime,
possibly accounting for the expected fact that the short-distance quantum structure
of spacetime would affect differently particles of different wavelengths (more strongly
the ones of shorter wavelength).
The notion of rainbow geometry
was introduced in \cite{mague04}, as a perspective on the structure of ``DSR-relativistic theories"\cite{AmelinoCamelia:2000mn,Magueijo:2002am, KowalskiGlikman:2002jr}, and since then it was widely used for quantum-gravity-phenomenology
analyses, see {\it e.g.} \cite{Smolin:2005cz, Garattini:2011hy, Ling:2005bp}.
While at first the notion of rainbow metric had only a semi-heuristic interpretation,
there are now formalisms that can consistently accommodate such a notion,
such as the relative-locality framework \cite{AmelinoCamelia:2011bm, AmelinoCamelia:2011pe, Gubitosi:2013rna, AmelinoCamelia:2011nt}
and certain Finsler geometries \cite{LiberatiFinsler06, kPoincareFinsler}.

We here propose to explore the possible role of rainbow metrics in singularity avoidance by following a full analysis
centered on the Raychaudhuri equation, an approach described in detail in the next section (Sec.\ \ref{rainbow}). This improves
on previous explorations of the same issue which relied on
a characterization of singularity avoidance based exclusively on the energy content of the universe through an extrapolation of the standard thermodynamics \cite{awad}. We thus provide an analysis of the implications of rainbow metrics on this issue which aims to account for all possible modifications.

Then in Sec.\ \ref{example}
we examine the quantitative implications of the improved methodology
of analysis here proposed for the cases of two particularly meaningful examples
of rainbow metrics.
This allows us to show that the improvements we advocate are not only valuable
conceptually but also have important  quantitative implications for the outcome of the analysis.

The closing Sec.\ \ref{conclusions} is devoted to a perspective on our proposals and results. There we also argue that our findings might suggest that  the semi-classical
approach which has been popular in the first explorations of rainbow cosmology, while being useful in getting hints about the early universe behavior,
might not be appropriate all the way back to the initial singularity, since in that regime quantum-gravity effects
not captured by the semi-classical approach may be relevant.

We use Planckian units such that $\hbar=K_B=c=1$. Notice that the deformations implied by MDRs and/or rainbow metrics
do not weaken the role of speed-of-light scale $c$ in physics, since that scale  still
is a relativistic invariant, the observer-independent value of the speed of massless particles in the infrared limit (the limit of small particle energy). For what concerns the speed of massless particles the only novelty one allows is that at high energies the speed might not be  $c$.

\section{Avoiding singularities in the rainbow universe}
\label{rainbow}

The existence of singularities in GR is an intrinsic feature of the theory, as seen through the
singularity theorems of Hawking and Penrose \cite{hawking}, which are based on the notion of geodesic
incompleteness. The analysis demands the use of the Raychaudhuri equation, i.e., the evolution equation
for the expansion coefficient $\theta\equiv V^\mu{}_{;\mu}$ of a congruence of curves defined by the
velocity field $V^\mu$. In the case of a congruence of geodesics it is given by
\begin{equation}
\dot\theta=2\omega^2-2\sigma^2-\frac{1}{3}\theta^2-R_{\mu\nu}V^\mu V^\nu, \label{eq:Rayeq}
\end{equation}
where dot means derivative with respect to an affine parameter, $2\omega^2\equiv
\omega_{\mu\nu}\omega^{\mu\nu}$ where $\omega_{\mu\nu}$ is the vorticity tensor and $2\sigma^2\equiv
\sigma_{\mu\nu}\sigma^{\mu\nu}$ where $\sigma_{\mu\nu}$ is the shear tensor associated to the
congruence. In the particular case of the Friedmann-Lema\^itre-Robertson-Walker (FLRW) model with velocity field
$V^\mu=\delta^\mu_0$, this equation reduces to the equation for the second derivative of the scale
factor. 

The important point here is that if $\dot{\theta}$ does not change sign, then inevitably the
congruence will shrink to a singularity if at some time $t_0$ we have $\theta_0<0$.
This can be easily seen as follows: suppose that it is possible to construct $V^\mu$ as a gradient, that is, we define a
congruence that is hypersurface orthogonal. In that case, the vorticity term $\omega^2$ vanishes. The
term related to the shear $\sigma^2$ doesn't change sign as it is a quantity orthogonal to $V^\mu$\footnote{We recall that a non-null shear can be a result of anisotropic expansion. Thus, as long as anisotropic spaces do not present vorticity, their singular behavior cannot be avoided based solely on the presence of shear.}. The
last term is proportional to $\rho +3p$ if GR is assumed. Therefore, if the strong energy condition
holds, that is $\rho+3p>0$, this term will never change sign as well.  It follows that \cite{hawking}
\begin{equation}
\frac{d\theta}{d\tau}\leq-\frac{1}{3}\theta^2\Rightarrow\theta^{-1}(\tau)\geq\frac{1}{3}\tau+\theta_0^{-1}.
\end{equation}

Thus, if $\theta_0<0$ for some value of  $\tau$, then $\theta^{-1}\rightarrow0^-$ and
$\theta\rightarrow-\infty$ within the proper time interval $\tau\leq3/|\theta_0|$.

In summary, upon assuming \begin{itemize}
\item the validity of GR and of the strong energy condition;
\item that gravity is strong enough to trap a region, so that one can have $\theta_0<0$ at some  $\tau_0$;
\item the existence of global hyperbolicity (which allows the definition of a global vector field with
    null vorticity);
\end{itemize}
one finds that the space-time is geodesically incomplete, that is, the history of a particle may have an end or a
beginning within a finite interval of proper time. These assumptions thus provide a set of conditions
that imply the existence of a singularity. More generally, different combinations of the conditions
above are possible. One could weaken one of them and assume stronger versions of the others to get
different versions of the theorem \cite{haw94}. If any of the conditions is not verified, then one can
not conclude anything about the existence of the singularities and should look directly at the
evolution equation, Eq. (\ref{eq:Rayeq}), and the invariants of the geometry constructed with it. Whereas previous studies of singularity avoidance in rainbow cosmology adopted simplified (and less conclusive) criteria, we shall here study geodesic (in)completeness through the Raychaudhuri equation (\ref{eq:Rayeq}).

In the context of a quantum-gravity theory, as mentioned, it is often useful to consider a semi-classical regime in
which specific quantum-gravitational effects can be effectively described in terms of modified
dispersion relations and the space-time can still be described as a smooth manifold, with non-trivial
properties (low-energy relics of quantum-gravitational effects) encoded exclusively in an effective
energy-dependent metric.

A generic MDR can be cast in the form
\begin{equation}
\label{MDR}
-f^2(E)E^2+g^2(E)p^2=m^2,
\end{equation}
where $f(E)$ and $g(E)$ are functions of the particle energy $E$ and of the deformation scale $E_P$,
which is expected to be of the order of the Planck energy. One must enforce of course  that $f(E)=1=g(E)$ in the limit $E
\ll E_P $, so that the standard special-relativistic dispersion relation is
recovered in that limit\footnote{Notice that this form of MDR is a  deformation of the special relativistic one, not
of the one one would have in a curved space-time. In this work we will assume that the generalization
to curved space-time MDR's can be made simply by taking into account redshift of energies, in the same
spirit of \cite{Jacob:2008bw}. A more rigorous
treatment would require to consider deformations of a proper dispersion relation  in curved space-time
and the associated rainbow metric, but unfortunately this kind of studies is still at a very
preliminary stage (see e.g. \cite{Marciano:2010gq}). }.

From a MDR of the form (\ref{MDR}) one can derive, in the case of a homogeneous and isotropic geometry
described by a FLRW-like metric, the following line element \cite{mague04}
\begin{equation}
\label{metric}
ds^2=-\frac{1}{f^2(E)}dt^2+\frac{a^2(t)}{g^2(E)}\delta_{ij}dx^idx^j.
\end{equation}
Here $a(t)$ is the standard expansion factor - the one describing the geometry probed by any particle
with energy $E\ll E_P$. When the particle energy is high enough that the modification of the dispersion relation, (\ref{MDR}), cannot be neglected, the particle will 'feel' the geometry described by
(\ref{metric}), with an effective expansion factor $a_{eff}\equiv a/g$ (see also discussion in
\cite{ling10}).

The metric (\ref{metric}) is explicitly time-dependent because of $a(t)$, but the time dependence comes
also indirectly from the functions $f$ and $g$. Indeed they depend on time through the energy, because
of the redshift due to space-time curvature. Considering the normalized velocity field
$V^\mu=f(E)\delta^\mu_0$ we thus obtain the expansion coefficient
 $$\theta=V^\mu{}_{;\mu}=3f\left(\frac{\dot a}{a}-\frac{\dot g}{g}\right)\equiv f\theta_{eff},$$
where $\theta_{eff}\equiv 3\dot a_{eff}/a_{eff}$ is defined in analogy with the FLRW case (where it
reads $\theta=3\dot{a}/a$) and whose evolution equation gives\footnote{This equation is not fully
equivalent to the one presented by \cite{ling10} and later by \cite{awad} because there is a missing
term in these references proportional to $\theta_{eff}^2$.}
\begin{equation}
\label{theta}
\dot\theta_{eff}+\frac{\theta^2_{eff}}{3}=-\frac{4\pi}{f^2}(\rho+3p)-\frac{\dot
f}{f}\frac{\theta_{eff}}{3},
\end{equation}
where dot from now on refers to derivative with respect to $t$. Before moving forward with the analysis it may be worth pausing here briefly to comment on our description of evolution in terms of the time variable $t$. At first sight one might be tempted to use a different time variable, say $t'$, obtained by absorbing in $t$ the energy-dependent function $f$. However, evidently such a map from $t$ to $t'$ would not be a diffeomorphism, indeed because of the energy dependence involved\footnote{Note that even though we have left the standard setting of Riemannian geometry, an energy-dependent metric can be accommodated for example in a Finsler geometry context \cite{LiberatiFinsler06, kPoincareFinsler} where the invariance under usual diffeomorphisms is unaltered and energy-dependent coordinate transformations are not in general symmetries of the geometry.}. Moreover, the new time variable $t'$ would not correspond to the readout of any physical clock: the energy dependence postulates in rainbow-gravity proposals is meaningful exclusively when the relevant energy scale is the energy of a fundamental particle, but one cannot assign a similar notion of energy to a physical clock. It is also well understood \cite{AmelinoCamelia:2011uk, Amelino-Camelia:2013fxa} that the relevant rainbow-gravity effects introduced for fundamental particles leave only negligible traces for macroscopic/composite bodies such as a physical clock. 

Note that if $f$ is a well behaved function - it should be bounded for large values of energy in order not to give a degenerate
metric and should smoothly tend to 1 in the low energy limit in order to recover the special relativistic dispersion
relation - the behavior of $\theta$ is reflected by the behavior of $\theta_{eff}$, and therefore we can
explicitly account for the contribution of $f$ to the evolution of the congruence by keeping it on the
right-hand side of the above equation.

The $0$-$0$ component of the Einstein equations reads
\begin{equation}
\label{fried}
\theta^2_{eff}=\frac{24\pi}{f^2}\rho,
\end{equation}
and the continuity equation is given by
\begin{equation}
\label{cont}
\dot\rho+\theta_{eff}(\rho+p)=0.
\end{equation}

Through Eq. (\ref{theta}) we see that there are basically two ways in which the initial singularity
could be avoided: either the condition $\rho+3p>0$ is violated due to modifications of the equation of
state $p=\omega(T)\rho$ - where the equation of state parameter becomes temperature-dependent - or the last term changes sign, which is possible depending on the form of
the function $f$.

The dependence on the temperature of the equation of state parameter arises from the modifications of the thermodynamics of the fluid being considered, as now the relation between energy and momentum is not trivial \cite{nozari, husain13,Alexander:2001,Alexander:2001ck}. The relevant thermodynamical quantities are computed using the density number of states with momenta in the interval between $p$ and $p+dp$ \cite{huang} which is defined by
\begin{equation}
\label{states}
N(p)dp=\frac{V}{\pi^2} p^2dp,
\end{equation}
and this relation gets modified when written in terms of energy due to the MDR, namely
\begin{equation}
N(E)dE=\frac{V}{\pi^2}\left(\frac{f}{g}\right)^3\left[1+\left(\frac{f'}{f}-\frac{g'}{g}\right)E\right]E^2dE,
\end{equation}
where $x'$ denotes derivative of $x$ with respect to $E$. Therefore, one has a modification of the total average energy, defined for a gas of photons by
\begin{equation}
U=\int E\frac{N(E)}{\exp{[E/T]}-1}dE,
\end{equation}
and consequently of the energy density $\rho=U/V$. Other quantities like pressure also get modified, implying a deviation of the equation of state parameter $\omega=p/\rho$ in the UV regime. In the next section we study two interesting cases where these modifications are not negligible.

\section{Quantitative analysis of some singular solutions}
\label{example}

In the previous section we have discussed what are the relevant quantities to be studied in order to
check for the presence/absence of singularities in a given rainbow universe.
In particular it is possible to check directly the evolution equation governing the behavior of the effective scale factor $a_{eff}$ which is given in
(\ref{theta}) and the equation of state parameter should be computed explicitly once we have the rainbow functions.

Before proceeding with our analysis, let us comment on some previous works dealing with the issue of
singularities in the rainbow universe, since the way they treat thermodynamics is significantly
different from what we do in this work, and leads to significantly different conclusions. The analysis
presented in \cite{awad} has assumed the classical equipartition theorem to relate the average energy
to the temperature without taking into account that a MDR should modify the thermodynamical properties
of the fluid. That would only be true if $f=g$ and for massless particles. Even if one wanted to claim
that the corrections would be of second order in the Planck length $l_{Pl}^2T^2$, and therefore
negligible \cite{ling06}, the specific form of the MDR is needed to justify it. This means that the
general criteria (i) and (ii) presented in \cite{awad} are not fully general, as the specific form of
the MDR should be used from the beginning. This argument applies also to the fluid equation of state,
which should also change due to the MDR. A proper analysis can be seen, for instance, in the context of
black holes \cite{amelino05}.

Besides, in Refs.\cite{ling06,ling10,awad} the authors offer an average
description of the metric (\ref{metric}), considered to be the effective metric probed by a radiation
fluid with average energy $E$, but it should be appreciated
that this leads to a picture in which the singularity is avoided on the average, while nothing can be said
about probes with energies different from the average one. We show in what follows with explicit
examples that this ambiguity is avoided when we consider a time dependence only through the actual
cosmological expansion $a(t)$, that is, the one defined for low values of energy, so that all the
probes exhibit the same time behavior even though they perceive different geometries.

\subsection{First example}
In order to clarify the procedure we suggest to follow, and in order to emphasize the possibly
different conclusions that we reach in comparison with previous analyses, let us now consider as an
example the MDR that was studied in \cite{awad}, where it was argued that it results in a non-singular
solution. This particular form of MDR is also connected rather directly to a quantum-gravity picture, whose interest was first stressed by
one of us in \cite{amelino98,AmelinoCamelia:2000mn}. It reads
\begin{equation}
\label{mdrex1}
\left(\frac{e^{E/E_{P}}-1}{E/E_{P}}\right)^2E^2=p^2,
\end{equation}
so that the rainbow functions are given by
\begin{equation}
\label{mdr1}
f(E)=\frac{e^{E/E_{P}}-1}{E/E_{P}}, \;\;\; g(E)=1.
\end{equation}
Firstly we study the thermodynamic properties of a gas of photons characterized by such a dispersion
relation, then we show how the modifications in the statistical quantities should be considered in
order to obtain explicitly the evolution of the scale factor, which turns out to be singular contrary
to what was claimed in \cite{awad}.

\subsubsection{Modified thermodynamics}
For an ensemble of particles, the density number of states with momenta in the interval between $p$ and
$p+dp$ is given by Eq. (\ref{states}). Therefore, it can be written in terms of energy by using the MDR governed by (\ref{mdr1}) as:
\begin{equation}
N(E)dE=\frac{V}{\pi^2} E^2_P(e^{E/E_P}-1)^2e^{E/E_P}dE.
\end{equation}

The equation of state parameter is defined in terms of the average energy density and the pressure
according to
\begin{equation}
\omega\equiv\frac{p}{\rho}=-T\fracc{\int \ln{[1-e^{-E/T}]}N(E)dE}{\int \frac{E}{\exp{[E/T]}-1}N(E)dE}.
\end{equation}

\begin{figure}[!htb]
\centering
\includegraphics[width=8cm,height=6cm]{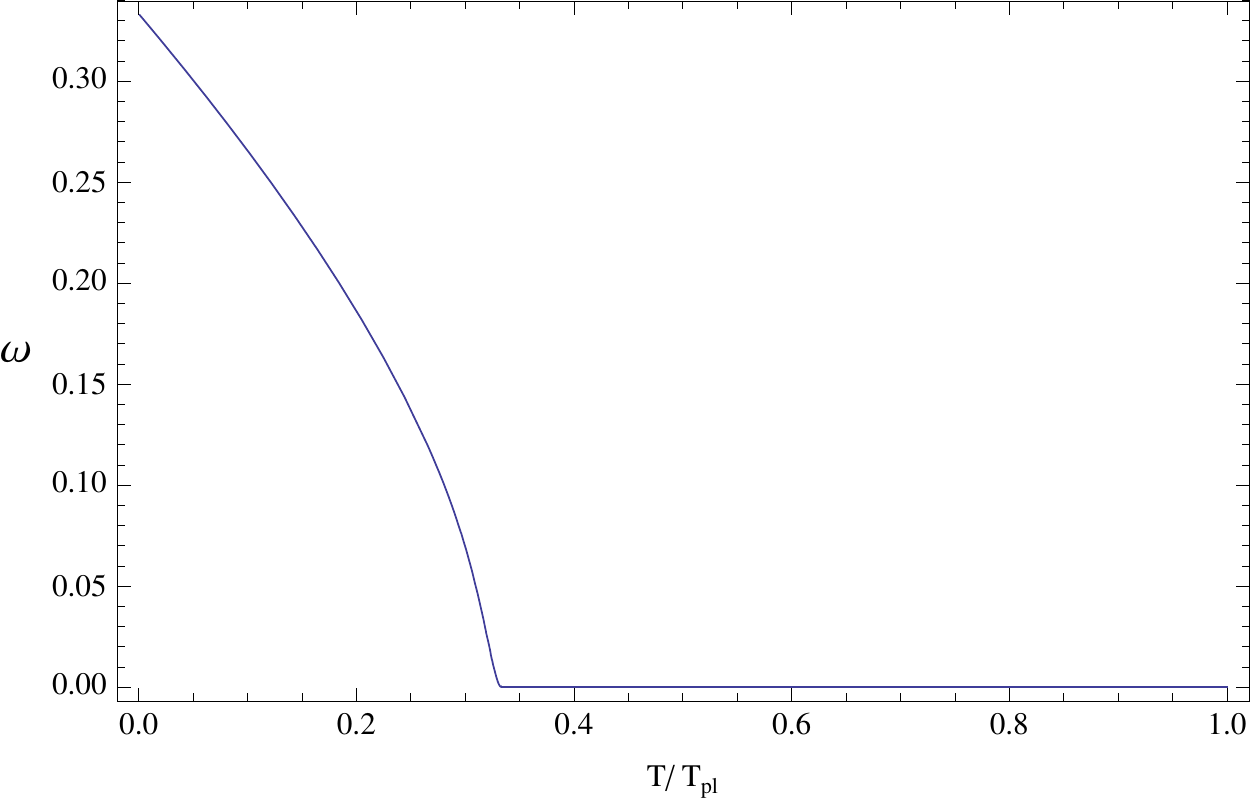}
\caption{Plot of the equation of state parameter as a function of temperature ($T_{Pl}$ is the Planck temperature).}
\label{eos1}
\end{figure}

The above expression can be numerically integrated to give us the equation of state parameter in terms
of temperature, as shown in Fig.\ref{eos1}.
We see that $\omega$ decreases from its usual value $1/3$ as the temperature increases and it quickly
approaches a constant asymptotic value, namely, $\omega\rightarrow 0$ in the UV regime. It is
interesting to note that in this regime the radiation fluid is effectively behaving as dust.

\subsubsection{Cosmological evolution and existence of singularity}
In this example, $g(E)=1$ and thus $a_{eff}$ coincides with the actual scale factor. Besides, according to what
has just been shown, we can assume that $\omega\approx 0$ in the UV regime. Then, the continuity
equation (\ref{cont}) gives us the energy density as a function of the scale factor as $\rho\propto
a^{-3}$. Besides, assuming that the time dependence of physical energy and momentum scales with the
scale factor as $p=k/a$, $k$ being the comoving wavenumber, we can obtain from (\ref{mdrex1}) the relation
\begin{equation}
E=E_P\ln{\left(\frac{k}{E_P a}+1\right)},
\end{equation}
so that the rainbow function $f(E)$ can be written as a function of the wavenumber $k$ (associated to
energy E) and of the scale factor:
\begin{equation}
f(a)=\fracc{k/(aE_P)}{\ln{\left(\frac{k}{E_P a}+1\right)}}.
\end{equation}

Plugging $\rho(a)$ and $f(a)$ into the first-order Friedmann equation, Eq. (\ref{fried}), we obtain
\begin{equation}\label{adot}
\dot{a}=\sqrt{\frac{8\pi\rho_{p}}{3}}\frac{E_P}{k}\sqrt{a}\ln{\left(\frac{k}{E_Pa}+1\right)},
\end{equation}
where $\rho_p$ is a constant of integration coming from the continuity equation. Note that this equation is valid only in the UV, for which the equation of state parameter takes the value $\omega=0$.

\begin{figure}[!htb]
\centering
\includegraphics[width=10cm,height=7cm]{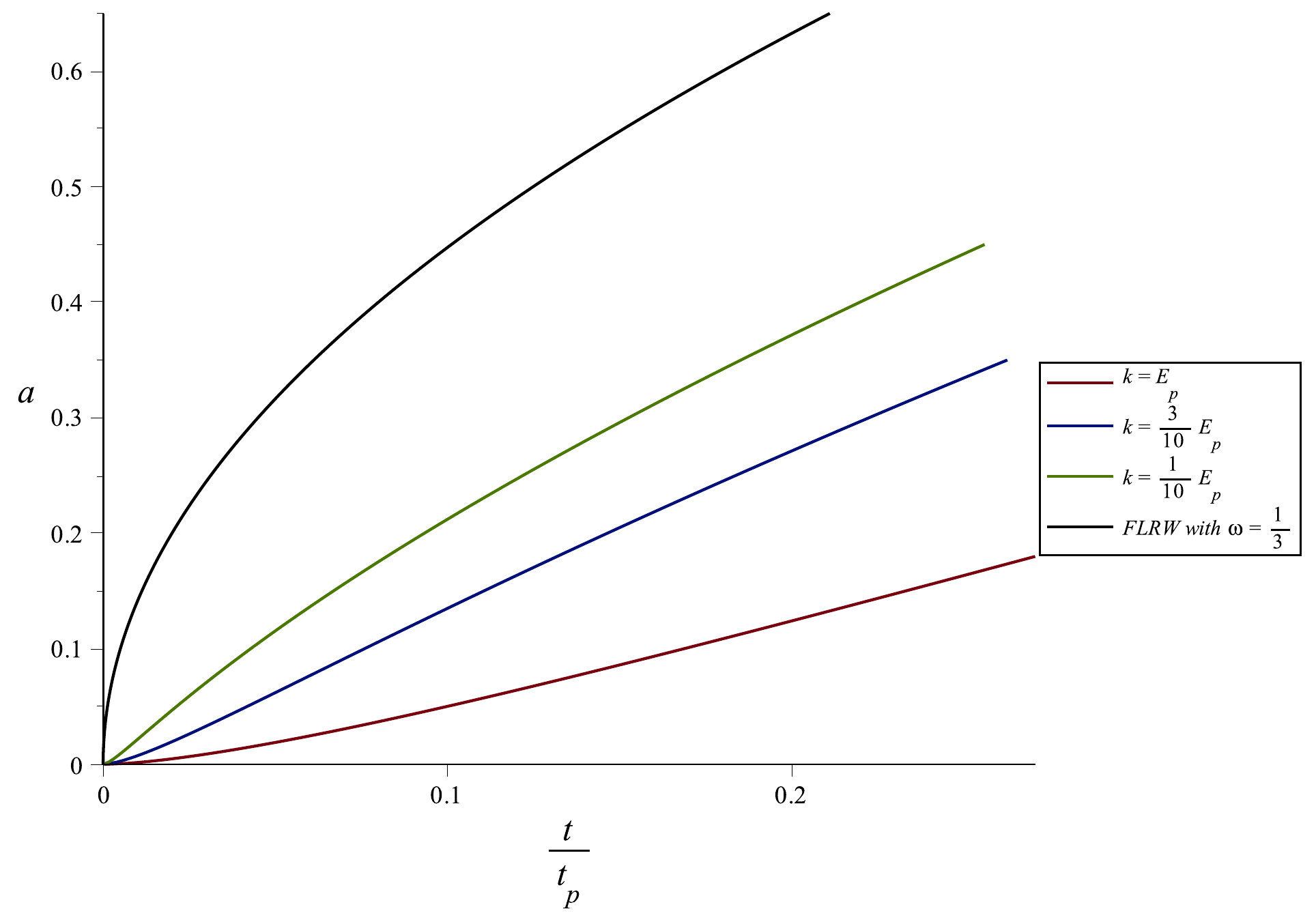}
\caption{{Comparison between different solutions for the scale factor where subscript $p$ refers to Planckian quantities. In black is plotted the standard radiation-dominated FLRW solution; in green, dark blue and red are, respectively, the solutions to the modified Eq. (\ref{adot}) for $k=1/10E_p$, $k=3/10 E_{P}$ and $k=E_P$. Note how the modified
evolution slows down as it approaches the singularity when compared to the standard FLRW
case. Here $t_p=1/\sqrt{8\pi\rho_p/3}$.}}
\label{scale}
\end{figure}

This equation leads to a singular solution as we can explicitly see in Fig.\ \ref{scale}, where we plot
the solution of this equation for different values of the wavenumber in comparison with the standard radiation-dominated FLRW solution.

We see that the modified evolution slows down near the singularity such that the higher the energy the
slower the approach. Indeed, the concavity of the curve changes for large values of energy indicating that quantum gravity effects are somehow smoothing the singularity. However, this is not enough to avoid the singularity, although it might be suggestive of the fact that
a full quantum-gravity picture, admitting our effective-theory description as an approximate description, could actually avoid the singularity. 

Another way of seeing this phenomenon of slowing-down is through the comparison between the Hubble parameter coming from Eq. (\ref{adot}) (defined as $H_k\equiv\dot{a}/a$) and the corresponding quantity in the FLRW radiation-dominated universe as the scale factor approaches zero. We remark that the Hubble parameter is invariant under spatial re-scaling transformations and therefore the slowing-down of the modified solutions is independent of the specific values we set for the comoving wavenumber. Thus, we have
\begin{equation}
\frac{H_{k}}{H}\propto a^{(3/2)}\ln{\left(\frac{k}{E_pa}+1\right)}\Rightarrow \lim_{a\rightarrow 0}\frac{H_{k}}{H} =0.
\end{equation}
That is, even though both quantities diverge as $a\rightarrow 0$, their ratio goes to zero for fixed values of $k$, meaning that the expansion rate for the FLRW case diverges faster than in the modified picture.

Note that in principle the singularity could be avoided from the arguments of section \ref{rainbow}.
The term $\dot{f}/f$ in Eq. (\ref{theta}) can be explicitly written as
\begin{equation}
\frac{\dot{f}}{f}=-\left(\frac{e^x}{e^x-1}-\frac{1}{x}\right)\left(\frac{1}{1+E_P a/k}\right)\frac{\theta_{eff}}{3},
\end{equation}
where $x\equiv E/E_P$. The term in the first brackets is always bigger than zero for all values of $x$.
Plugging this expression back into Eq. (\ref{theta}) it is possible to see that it contributes with an
opposite sign and thus, in principle, it could prevent the divergence of $\theta_{eff}$. The fact that our semi-classical description, based on an ``effective" rainbow metric, is unreliable as the scale factor goes to zero is seen through the fact
that arbitrarily high values of energy can be achieved and therefore the rainbow function and consequently $\theta_{eff}$ diverge.\footnote{It could be the case that a MDR is capable of providing a bouncing solution such
that the bounce would occur at an energy scale several orders of magnitude smaller than the Planck
energy. Then an analysis based on the full quantum gravitational theory wouldn't be strictly
necessary.}

\subsection{Second example}
Let us now consider another much studied example of the MDR for massless particles, given by the relation
\begin{equation}
\label{mdr}
E^2=p^2[1+(\lambda p)^{4}],
\end{equation}
so that the rainbow functions are $f=1$ and $g(p)=[1+(\lambda p)^{4}]^{1/2}$. $\lambda$ is a deformation parameter with dimensions of inverse of energy. This MDR is found to
provide an invariant power spectrum for cosmological perturbations and also gives the expected
spectral dimension reduction found in different approaches to quantum gravity \cite{qgdr}: the UV value
of the spectral dimension is two. Again, we first analyze its thermodynamic properties and then we
study this fluid in the modified FLRW geometry (\ref{metric}), showing that the cosmological evolution
also contains a singularity.

\subsubsection{Modified thermodynamics}

\begin{figure}[!htb]
\centering
\includegraphics[width=8cm,height=6cm]{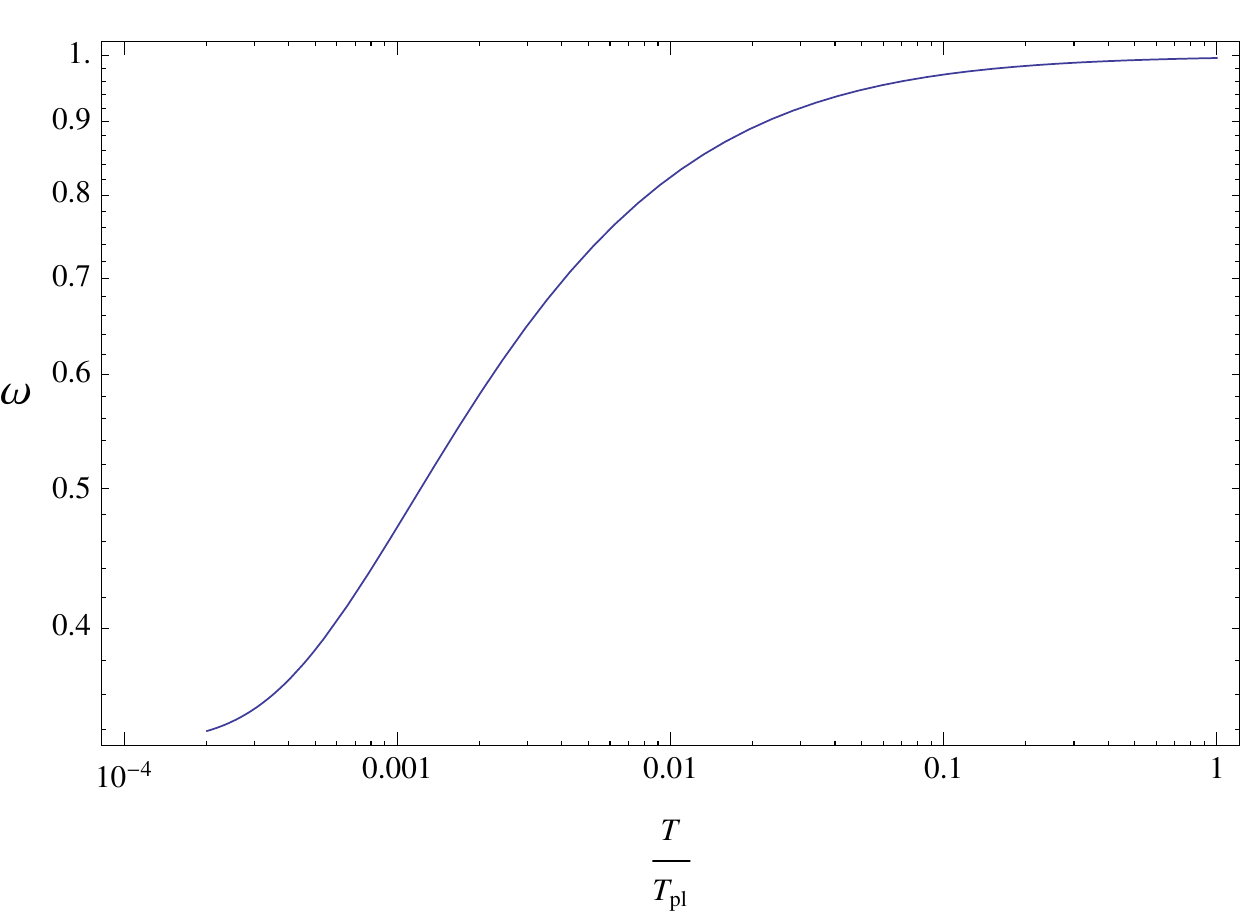}
\caption{Logarithmic plot of the equation of state parameter as a function of $T/T_P$ with
$\lambda=10^{5}L_P$, with $L_P$ the Planck length.}
\label{eosgraph}
\end{figure}

The density of states is given as before by Eq. (\ref{states}) where we now keep the dependence on $p$.
As in the first example, the equation of state parameter varies, departing from the usual value $1/3$
and approaching $\omega=1$ in the UV limit, as seen from the numerical integration of the expression
\begin{equation}
\omega=\frac{p}{\rho}=-T\fracc{\int \ln{[1-e^{-E(p)/T}]}N(p)dp}{\int
\frac{E(p)}{\exp{[E(p)/T]}-1}N(p)dp},
\end{equation}
in Fig. \ref{eosgraph}.

\begin{figure}[!htb]
\centering
\includegraphics[width=8cm,height=6cm]{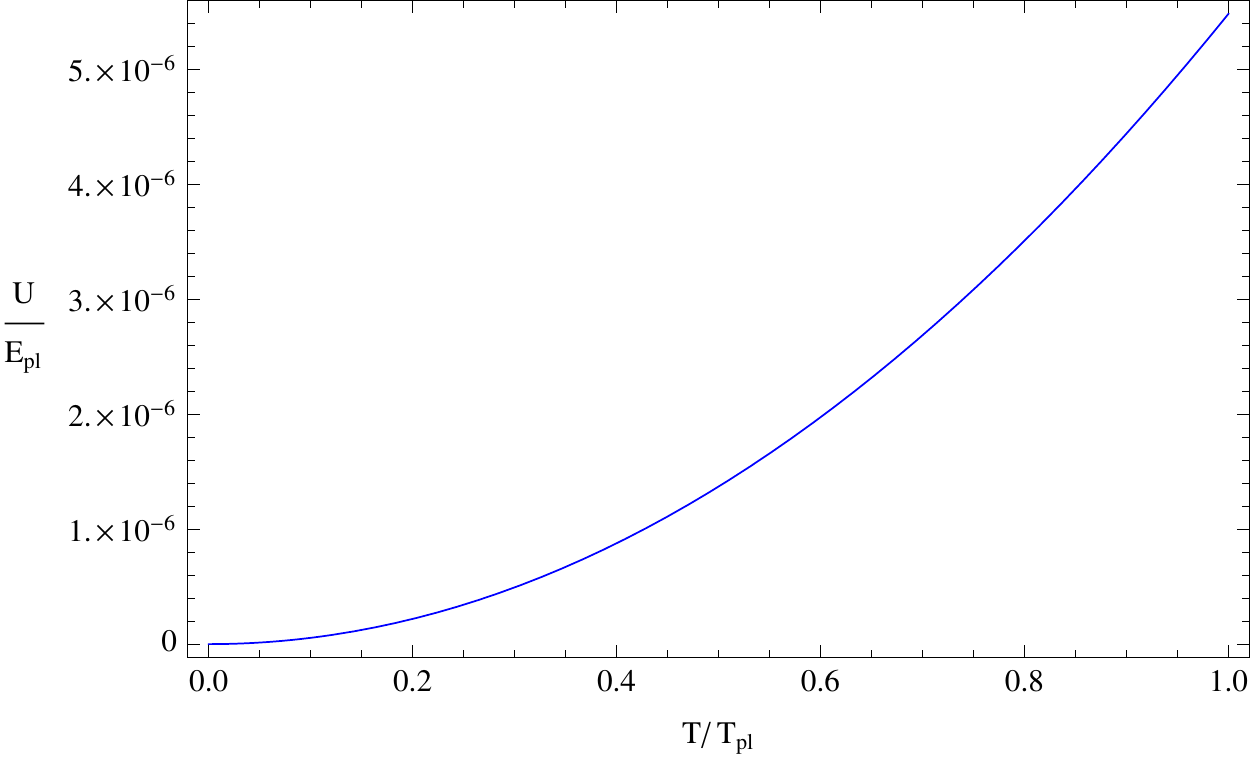}
\caption{Average energy as a function of $T$ with $\lambda=10^{5}L_P$.}
\label{energygraph}
\end{figure}

It is intriguing to observe that the UV value of the equation of state parameter matches the one of a
radiation fluid in two spacetime dimensions, which is also the value of the UV spectral dimension that can be
deduced from the dispersion relation (\ref{mdr}) \cite{amelino13}. Moreover, also the energy density
behaves in the UV as if the radiation fluid was two-dimensional. To show this we perform a numerical
integration and find the total energy as a function of temperature, through the expression
\begin{equation}
\label{energy}
U=\int E(p)\frac{N(p)}{\exp{[E(p)/T]}-1}dp,
\end{equation}
which yields the curve shown in Fig. \ref{energygraph}.

It is possible to see that the system seems to behave as a conventional one (usual dispersion
relation), but in two space-time dimensions. By fitting the curve one can easily check that in the UV it goes
like $T^2$. If we assume that the total energy of the system goes with $U\propto T^{d+1}$, where $d$ is
the number of spatial dimensions \cite{nozari, husain13}, we see that our system is effectively
behaving like one with the usual dispersion relation in one spatial dimension.

\subsubsection{Cosmological evolution and existence of singularity}
Again, the time dependence of the function $g$ comes only from the time dependence of the actual scale
factor $a(t)$, thus it is the same for all particles with different momenta. This means that even if
the probes are capable of perceiving only the effective scale factor $a_{eff}$, and the latter is
dependent on the wavenumber of the probe, the presence of a singularity will be a general feature of
the space-time.

\begin{figure}[!htb]
\centering
\includegraphics[width=9cm,height=7cm]{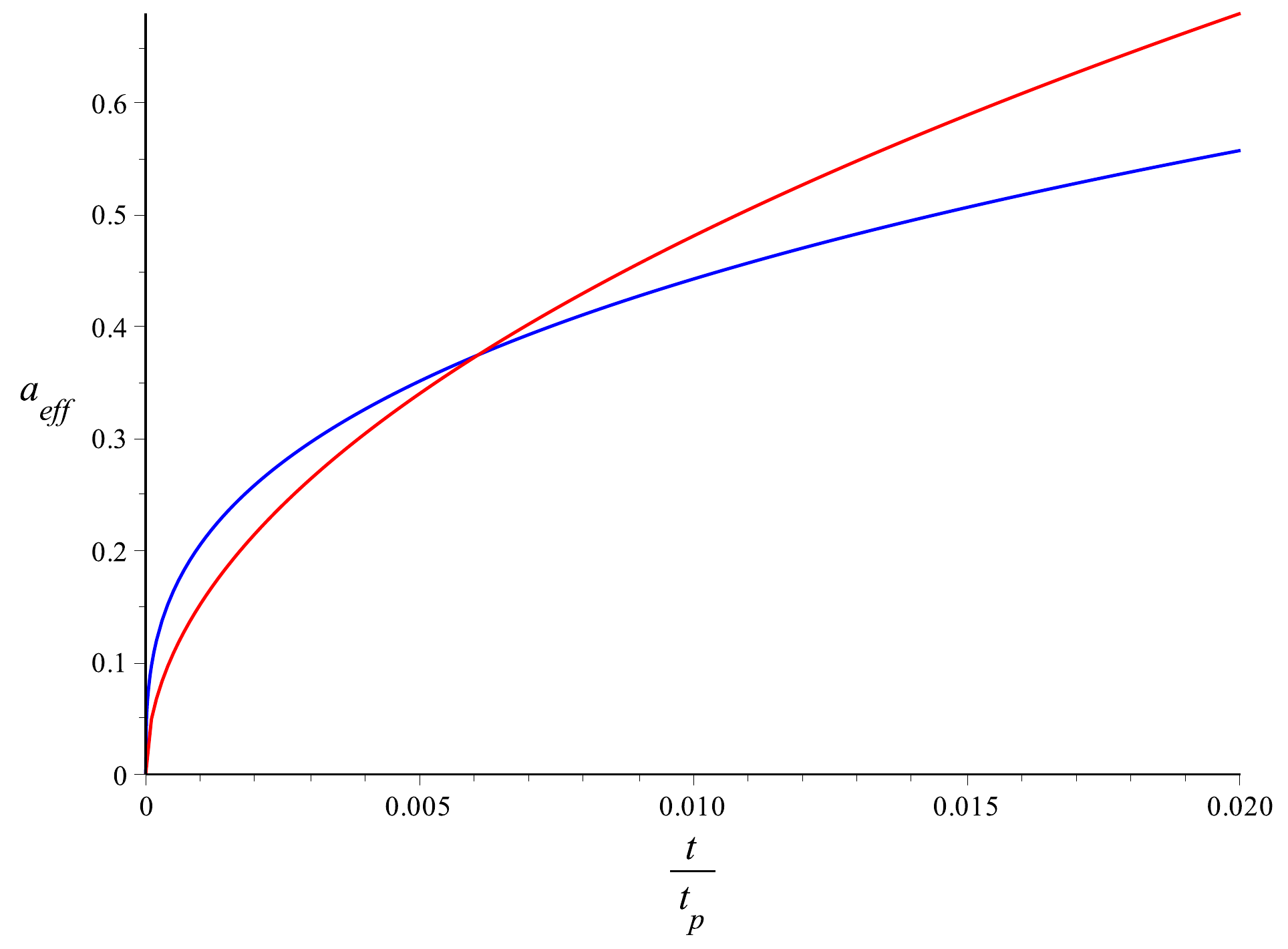}
\caption{Comparison between the standard evolution of the scale factor in a radiation dominated FLRW
universe (in red) for which $g(E)=1$ and the evolution of the effective scale factor in the case of the MDR given in Eq. (\ref{mdr})
(in blue) for a fixed value of the wavenumber.}
\label{2nd}
\end{figure}

In the case we are considering $f=1$, the only possibility to avoid a singularity according to
Eq.(\ref{theta}) would be a non-conventional behavior of the fluid such that it could violate the
energy condition $\rho+3p>0$. As we saw in the last subsection, this is not the case for the MDR
considered, as the equation of state parameter asymptotically goes to the constant value $\omega\approx
1$, resulting in a time dependence of the form $a_{eff}(t)\propto t^{1/3}$. Therefore, this rainbow
universe is singular, as can be explicitly seen in Fig. \ref{2nd}.

We see that the modified evolution seems to be slower when compared to the radiation dominated FLRW case for most of the evolution of the universe, which
is an indication that it could bypass the singular behavior. However, as we follow it backwards in time it
actually reaches the singularity even more abruptly than in the standard case. Again, the analysis indicates
that the full quantum gravity should be applied in the far UV regime, where the semi-classical picture is no longer an accurate description of the quantum universe.

\section{Conclusions}
\label{conclusions}
As some previous studies we here worked within the legitimate assumption (legitimate but only one of many possible hypotheses) that a rainbow metric (\ref{metric}) could be used for a reliable approximate effective-theory description of some quantum-gravity effects. Our strategy of analysis provides an improvement on previous analogous studies in several ways. We have shown (and used the fact) that a proper
analysis of the possible avoidance of the initial singularity should take into account the modified
thermodynamics implied by the MDR, as it can drastically change the time behavior of the scale factor.
Our approach is also not based on an average description as the one introduced in
\cite{ling06} and later used in \cite{awad}, so it is more conclusive in that our findings apply also to  probes with energy different from the average one. We have argued that it is possible
to provide a consistent analysis of the issue considering a time dependence that comes only from the
cosmological expansion. We have provided explicit computations showing, in two different pictures, that our concerns are not merely conceptual but rather have significant quantitative implications (to the point that pictures found to avoid the singularity in other more simplified approaches are not found to avoid the singularity within our approach). Still, from the physics perspective the most exciting feature is the one we found in the first of the two specific models here analyzed, where the approach to the singularity is slowed down by the MDR. As also supported by some of the points made here, such cases where singularity approach is slowed down within the effective-theory description based on a rainbow metric may well ultimately be part of a quantum-gravity picture in which the singularity is avoided, even though extrapolating the rainbow-metric approach to very small scale factors we do still find the singularity. This is plausible because the onset of features slowing down the singularity approach occurs in a regime which may well be within the reach of the effective-theory description based on rainbow metrics, whereas at even earlier times (smaller values of the scale factor) the reliability of this effective description most evidently  breaks down. The additional quantum-gravity effects that one should take into account at that stage, intervening in a scenario where singularity approach has already been slowed down, might well provide the needed additional structures for avoiding the singularity. 
\begin{acknowledgments}
GAC and GG are supported by grants from the John Templeton Foundation. 
GS acknowledges the support by the CAPES-ICRANet program financed by CAPES - Brazilian Federal Agency
for Support and Evaluation of Graduate Education within the Ministry of Education of Brazil through the
grant BEX 13955/13-6.
\end{acknowledgments}


\begin{thebibliography}{20}

\bibitem{novello}
M. Novello and S. E. P. Bergliaffa, {\em Phys. Rept.} {\bf 463}, 127 (2008) [arXiv:0802.1634
[astro-ph]].

\bibitem{asht11}
A. Ashtekar and P. Singh, {\em Class. Quant. Grav.} {\bf 28}, 213001 (2011) [arXiv:1108.0893 [gr-qc]];
N. Pinto-Neto and J. C. Fabris, {\em Class. Quant. Grav.} {\bf 30}, 143001 (2013) [arXiv:1306.0820
[gr-qc]]; L. J. Garay, M. Martin-Benito and E. Martin-Martinez, {\em Phys. Rev. D} {\bf 89}, 043510
(2014) [arXiv:1308.4348 [gr-qc]]; G. Calcagni, JHEP 0909:112 (2009) [arXiv:0904.0825 [hep-th]]; R.
Brandenberger, {\em Phys. Rev. D} {\bf 80}, 043516 (2009) [arXiv:0904:2835 [hep-th]]; D. Battefeld and
P. Peter, {\em Phys.\ Rept.} {\bf 571} (2015) 1 [arXiv:1406.2790 [astro-ph.CO]].

\bibitem{AmelinoCamelia:1997gz}
  G.~Amelino-Camelia, J.~R.~Ellis, N.~E.~Mavromatos, D.~V.~Nanopoulos and S.~Sarkar,
  {\em Nature} {\bf 393} (1998) 763
  [astro-ph/9712103].

\bibitem{Gambini:1998it}
  R.~Gambini and J.~Pullin,
  {\em Phys.\ Rev.\ D} {\bf 59} (1999) 124021
  [gr-qc/9809038].

\bibitem{Alfaro:1999wd}
  J.~Alfaro, H.~A.~Morales-Tecotl and L.~F.~Urrutia,
  {\em Phys.\ Rev.\ Lett.}\  {\bf 84} (2000) 2318
  [gr-qc/9909079].

\bibitem{AmelinoCamelia:1999pm}
  G.~Amelino-Camelia and S.~Majid,
  {\em Int.\ J.\ Mod.\ Phys.\ A} {\bf 15} (2000) 4301
  [hep-th/9907110].


\bibitem{mague04}
 J.~Magueijo and L.~Smolin,
  {\em Class.\ Quant.\ Grav.}\  {\bf 21} (2004) 1725
  [arXiv:0305055 [gr-qc]].


\bibitem{AmelinoCamelia:2000mn}
  G.~Amelino-Camelia,
  {\em Int.\ J.\ Mod.\ Phys.\ D} {\bf 11} (2002) 35
  [gr-qc/0012051].



\bibitem{KowalskiGlikman:2002jr}
  J.~Kowalski-Glikman and S.~Nowak,
  {\em Int.\ J.\ Mod.\ Phys.\ D}\ {\bf 12} (2003) 299
  [hep-th/0204245].


\bibitem{Magueijo:2002am}
  J.~Magueijo and L.~Smolin,
  {\em Phys.\ Rev.\ D} {\bf 67} (2003) 044017
  [gr-qc/0207085].




\bibitem{Smolin:2005cz}
  L.~Smolin,
  {\em Nucl.\ Phys.\ B} {\bf 742} (2006) 142
  [arXiv:0501091 [hep-th]].

\bibitem{Garattini:2011hy}
  R.~Garattini and G.~Mandanici,
  {\em Phys.\ Rev.\ D} {\bf 85} (2012) 023507
  [arXiv:1109.6563 [gr-qc]].

 \bibitem{Ling:2005bp}
  Y.~Ling, X.~Li and H.~b.~Zhang,
  {\em Mod.\ Phys.\ Lett.\ A} {\bf 22} (2007) 2749
  [arXiv:0512084 [gr-qc]].

\bibitem{AmelinoCamelia:2011bm}
  G.~Amelino-Camelia, L.~Freidel, J.~Kowalski-Glikman and L.~Smolin,
  {\em Phys.\ Rev.\ D} {\bf 84} (2011) 084010
  [arXiv:1101.0931 [hep-th]].

\bibitem{AmelinoCamelia:2011pe}
  G.~Amelino-Camelia, L.~Freidel, J.~Kowalski-Glikman and L.~Smolin,
  {\em Gen.\ Rel.\ Grav.}\  {\bf 43} (2011) 2547
  [arXiv:1106.0313 [hep-th]].

\bibitem{Gubitosi:2013rna}
  G.~Gubitosi and F.~Mercati,
  {\em Class.\ Quant.\ Grav.}\  {\bf 30} (2013) 145002
  [arXiv:1106.5710 [gr-qc]].

\bibitem{AmelinoCamelia:2011nt}
  G.~Amelino-Camelia, M.~Arzano, J.~Kowalski-Glikman, G.~Rosati and G.~Trevisan,
  {\em Class.\ Quant.\ Grav.}\  {\bf 29} (2012) 075007
  [arXiv:1107.1724 [hep-th]].


  \bibitem{LiberatiFinsler06}
  F.~Girelli, S.~Liberati and L.~Sindoni,
  {\em Phys.\ Rev.\ D} {\bf 75} (2007) 064015
  [arXiv:0611024 [gr-qc]].

\bibitem{kPoincareFinsler}
  G.~Amelino-Camelia, L.~Barcaroli, G.~Gubitosi, S.~Liberati and N.~Loret,
  [arXiv:1407.8143 [gr-qc]].

\bibitem{awad}
A. Awad, A. F. Ali and B. Majumder, JCAP {\bf 10}, 052 (2013) [arXiv:1308.4343 [gr-qc]].

\bibitem{hawking}
S. W. Hawking and G. F. R. Ellis, {\em The large scale structure of space-time}, Cambridge University
Press (1973).

\bibitem{haw94}
S. W. Hawking, arXiv:9409195 [hep-th].

\bibitem{Jacob:2008bw}
M. R. Martinez and T. Piran, JCAP {\bf 0604} (2006) 006 [arXiv:0601219[astro-ph]];
  U.~Jacob and T.~Piran,
  JCAP {\bf 0801} (2008) 031
  [arXiv:0712.2170 [astro-ph]].

\bibitem{Marciano:2010gq}
  A.~Marciano, G.~Amelino-Camelia, N.~R.~Bruno, G.~Gubitosi, G.~Mandanici and A.~Melchiorri,
  JCAP {\bf 1006} (2010) 030
  [arXiv:1004.1110 [gr-qc]].


\bibitem{ling10}
Y. Ling and W. Wu, {\em Phys. Lett. B} {\bf 687}, 103 (2010) [arXiv:0811.2615v2 [gr-qc]].

\bibitem{AmelinoCamelia:2011uk}
  G.~Amelino-Camelia, L.~Freidel, J.~Kowalski-Glikman and L.~Smolin,
  {\em Phys.\ Rev.\ D} {\bf 84} (2011) 087702
  [arXiv:1104.2019 [hep-th]].

\bibitem{Amelino-Camelia:2013fxa}
  G.~Amelino-Camelia,
  {\em Phys.\ Rev.\ Lett.}\  {\bf 111} (2013) 101301
  [arXiv:1304.7271 [gr-qc]].

\bibitem{nozari}
K. Nozari and A.S. Sefidgar, {\em Chaos, Solitons and Fractals} {\bf 38}, 339 (2008) .

\bibitem{husain13}
V. Husain, S. S. Seahra and E. J. Webster, {\em Phys. Rev. D} {\bf 88}, 024014 (2013) [arXiv:1305.2814
[hep-th]].

\bibitem{Alexander:2001}
  S.~Alexander, R.~Brandenberger and J.~Magueijo,
  {\em Phys.\ Rev.\ D }{\bf 67} (2003) 081301
  [hep-th/0108190].
\bibitem{Alexander:2001ck}
  S.~Alexander and J.~Magueijo,
  Proceedings of the XIIIrd Rencontres de Blois 'Frontiers of the Universe', pp281, The Gioi Publishers, 2004
  [hep-th/0104093].

\bibitem{huang}
See, for example: K. Huang, {\em Statistical Mechanics}, 2nd edition, John Wiley \& Sons, New York (1987).

\bibitem{ling06}
Y. Ling, JCAP {\bf 08}, 017 (2007) [arXiv:0609129 [gr-qc]].

\bibitem{amelino05}
G. Amelino-Camelia, M. Arzano, Y. Ling and G. Mandanici, {\em Class. Quant. Grav.} {\bf 23}, 2585
(2006) [arXiv:0506110 [gr-qc]].

\bibitem{amelino98}
G. Amelino-Camelia, J. R. Ellis, N. Mavromatos, D. V. Nanopoulos and S. Sarkar, {\em Nature} {\bf 393},
763 (1998) [arXiv:9712103 [astro-ph]]; G. Amelino-Camelia, {\em New J. Phys.} {\bf 6}, 188 (2004).

\bibitem{qgdr}
J. Ambjorn, J. Jurkiewicz and R. Loll, {\em Phys. Rev. Lett} {\bf 95}, 171301 (2005) [arXiv:0505113
[hep-th]]; P. Horava, {\em Phys. Rev. Lett} {\bf 102}, 161301 (2009) [arXiv:0902.3657 [hep-th]]; D. F.
Litim, {\em Phys. Rev. Lett} {\bf 92}, 201301 (2004) [arXiv:0312114 [hep-th]]; L. Modesto, {\em Class.
Quant. Grav.} {\bf 26}, 242002 (2009) [arXiv:0812.2214 [gr-qc]];

\bibitem{amelino13}
G. Amelino-Camelia, M. Arzano, G. Gubitosi and J. Magueijo, {\em Phys. Rev. D} {\bf 87}, 123532 (2013)
[arXiv:1305.3153 [gr-qc]] and {\em Phys. Rev. D} {\bf 88}, 041303 (2013) [arXiv:1307.0745 [gr-qc]].



\end{thebibliography}
\end{document}